# An equilibrium model for RFP plasmas in the presence of resonant tearing modes

P. Zanca[(*)], F. Sattin

*Consorzio RFX, ENEA-Euratom association*
*Corso Stati Uniti 4, 35127 Padova*
*ITALY*

**Abstract**

The equilibrium of a finite-$\beta$ RFP plasma in the presence of saturated-amplitude stationary tearing modes is investigated. The singularities of the MHD force balance equation $\mathbf{J} \times \mathbf{B} = \nabla p$ at the modes rational surfaces are resolved through a proper regularization of the cylindrical symmetric component of the profiles, by setting to zero there the gradient of the pressure and parallel current density. An equilibrium model, which satisfies the regularization rule at the various rational surfaces, is developed. The comparison with the experimental data from the Reversed Field eXperiment (RFX) gives encouraging results. We will show that the model provides an easy tool for magnetic analysis.

## 1. Introduction

In a Reversed Field Pinch (RFP) plasma the maintenance of the configuration against resistive diffusion is provided by the plasma itself through the generation of magnetohydrodinamical (MHD) modes (dynamo modes) [1]. RFX, a high-current RFP [2], exhibits a strong MHD dynamics with many saturated-amplitude stationary modes. As a consequence, the RFX equilibrium configuration is non-axisymmetric and essentially three-dimensional (3-D). However, for an RFP the dominant part of any spatially dependent field is likely to be still cylindrically symmetric, and the deviation from axisymmetry can be regarded as a perturbation of this zeroth-order component. Therefore, any given field $A$ can be written as:

1) $\quad A(r,\theta,\phi) = A_0(r) + a(r,\theta,\phi); \quad a(r,\theta,\phi) = \sum_{m,n} a^{m,n}(r) e^{i(m\theta - n\phi)}; \quad |a^{m,n}| << |A_0|;$

We choose to work in a cylindrical geometry, hence the system is taken to be periodic in the $z$-direction with periodicity length $2\pi R$, and a simulated toroidal angle $\phi = z/R$ is adopted. The Fourier

---

[(*)] E-mail: zanca@igi.pd.cnr.it

harmonics related to the modes are characterized by a poloidal mode number *m,* and a toroidal mode number *n (n≠0)*. The most important instabilities thought to exist in an RFP are tearing modes [3, 4]. They are resonant modes, i.e. a flux-surface for which the safety factor satisfies the relation $q(r = r_s^{m,n}) = m/n$, exists inside the plasma. The flux-surface where the previous condition holds is the *(m, n)* rational surface. Therefore in the expansion (1) we will take into consideration only resonant harmonics. Tearing modes develop owing to the small but finite resistivity of the plasma. The resistivity breaks the flux-freezing constraint in the Faraday-Ohm's law, and allows the tearing and reconnection of equilibrium magnetic flux surfaces, to produce helical magnetic islands centred on each rational surface.

The purpose of this paper is to study the RFP equilibrium magnetic configuration for fields defined by (1) using the MHD force balance equation

2) $\mathbf{J} \times \mathbf{B} = \nabla p$

together with the Ampere's law and the divergence law.

3) $\nabla \times \mathbf{B} = \mu_0 \mathbf{J} \quad ; \quad \nabla \cdot \mathbf{B} = 0$

We have set $\rho \partial \mathbf{v}/\partial t = 0$ in (2) in order to study a configuration in which the modes are stationary both in amplitude and in phase. This is in fact the standard situation in RFX. Moreover, the convective inertial term $\rho \mathbf{v} \cdot \nabla \mathbf{v}$ and the viscous force $\nu \nabla^2 \mathbf{v}$ have been neglected, because they are typically small effects (this fact will be discussed again below). Therefore our system of equations (2, 3) is closed and we don't need to introduce the Faraday-Ohm's law. Nevertheless, the effect of resistivity in the Faraday-Ohm's equation is implicitly taken into account because we leave the possibility for having non-zero radial field harmonics at the various rational surfaces.

It is known that the MHD equilibrium of a 3-D configuration is characterized by the presence of singularities at the rational magnetic surfaces [5, 6]. For example, in the stellarator configuration the parallel component of the pressure driven current (Pfirsch-Schlütler current) can develop singular terms at the rational surfaces [7]. If a rational surfaces exhibits such infinities, it is possible to adopt a regularization of the equilibrium fields in order to avoid them [5, 6].

Despite the difference between the RFP and stellarator configurations we will show that the singularities of equation (2) in the presence of the resonant harmonics (1) can be healed by a regularization of the zeroth-order profiles which is similar to those described in [5, 6]. It will be discussed later that this regularization imposes the vanishing of the gradients of the zeroth-order pressure and parallel current at the modes rational surfaces. The justification of the regularization procedure relies on the fact that we have indeed a large latitude for the choice of the zeroth-order profiles. Moreover, it is a well known fact that tearing islands do imply a local flattening of both pressure and current profile [8]. However, we note that the dynamical growing phase of the modes is led by the pressure and current gradients and it is described by the standard layer/ island analysis [9], where the singularity is in this case resolved by considering a different time-dependent equation in a narrow region about the rational surface. Therefore our regularization rule does apply to the saturated-amplitude stationary phase only. Note that the regularization establishes a feedback effect of saturated instabilities on the cylindrical-symmetric zeroth-order profiles. In fact, in the presence of many resonant modes the combined effect of the regularization at the different rational surfaces determines an overall flattening of the profiles: the more modes we have the flatter the profile we get. We want to point out that our approach is just complementary to a stability analysis: while the latter starts from a given equilibrium profile and determines the growth rates of the modes, our analysis attempts to describe the feedback effect of saturated modes over the zeroth-order profile.

Tokamak plasmas are almost totally driven by external means and it is likely that the influence of instabilities on equilibrium profile be small here. On the other hand, in RFPs plasmas the dynamo modes determine a relaxation (flattening) of the equilibrium profiles which drive the configuration near the Taylor's minimum energy state [10]. Therefore the regularization procedure seems a suitable way to quantify the relaxation of the zeroth-order profiles due to these modes. There are mainly two experimental observations that support this hypothesis: a) in RFX several resonant instabilities are observed and the parallel current profile is <u>very flat</u>. This is not only indicated by the standard $\alpha$-$\Theta_0$ equilibrium model [1], commonly adopted for RFPs, but it is also confirmed and reinforced by the internal polarimetric measurements [11]. b) When some of the secondary $m=1$ modes disappear, either for a spontaneous process ($\alpha$-mode [12]) or for an external influence (PPCD [13]), a <u>steepening</u> of the parallel current profile is observed. The same behaviour is observed for the pressure, even if only the electronic component can be measured [14]. Another confirmation of our ideas comes from the recently published reconstruction of the MST equilibrium profiles [15].

The regularization rule brings in a natural way to a numerical model for computing, under some reasonable assumptions, the zeroth-order profiles. The required informations are the global equilibrium parameters $\Theta$, $F$, $\beta_p$ and the number of observed instabilities. Given the zeroth-order profile, the various perturbed harmonics can be determined in a straightforward way. Therefore this approach provides a tool for magnetic analysis that, in our opinion, is worthy of being investigated.

In principle the equilibrium system (2, 3) could be improved by adding the convective inertial term and the viscous term in the force balance equation (2). In this way the steady-state Faraday-Ohms's law must be introduced in order to close the equations. This system is analyzed using a simplified slab geometry in the Appendix A. We think that the regularization rule is not significantly changed in this new system. In fact the scale length over which the velocity perturbation varies near the rational surfaces is not likely to be smaller than the width of the islands associated to the modes. Under this hypothesis, for sufficiently high mode amplitudes (like those observed in the RFX experiment) the extra inertia and viscous terms give a very small contribution also in the island region. This seems to be confirmed by the fact that in RFX the tearing modes are always observed to be phase-locked: according to a recent analysis [16] the phase-locking is the condition under which the electromagnetic torque developed at the rational surfaces by the non-linear interaction between the modes has the minimum amplitude; the need of having a phase-locking in the steady-state motion equation means that at the rational surfaces the inertia and viscous terms are sub-dominant with respect to the electromagnetic term. In conclusion, we think that the equilibrium system of equations (2, 3) represents a good approximation to describe a high mode amplitude regime.

The plan of the work is the following: in Section 2 we analyze the problem of the equilibrium in the presence of saturated resonant instabilities, showing how to overcome the singularity of the force balance equation by an ad-hoc regularization of zeroth-order profiles. Sections 3 enters in the details of the regularization method, showing practically how it is expected to modify equilibrium profiles. Section 4 is devoted to the description of the numerical algorithm adopted for solving equilibrium equations. Section 5 gives examples of application of the algorithm to specific RFX plasmas: zeroth-order profiles derived by our model are computed and compared with the standard $\alpha$-$\Theta_0$ prediction. We also give an estimate for perturbation profiles and their amplitudes inside the plasma, starting from the measurement of their values at the edge. This allows the spatial reconstruction of the total field perturbation. To this regard, we point out that this work may be seen as a continuation and a completion of a previous analysis about the magnetic perturbation in the vacuum region of RFX [17, 18]. Conclusions are drawn in section 6.

### 2. Equilibrium equations
*2.1 Zeroth-order fields*

The components of the zeroth-order fields are

4) $\mathbf{B}_0 = [0, B_{0\theta}(r), B_{0\phi}(r)]; \quad \mathbf{J}_0 = [0, J_{0\theta}(r), J_{0\phi}(r)]$

From (2, 3) one gets

5) $\nabla \times \mathbf{B}_0 = \mu_0 \dfrac{\mathbf{J}_0 \cdot \mathbf{B}_0}{B_0^2} \mathbf{B}_0 - \mu_0 \dfrac{\nabla p_0 \times \mathbf{B}_0}{B_0^2}$

Defining the normalized pressure gradient and the parallel current density profile as

6) $g(r) = \dfrac{\mu_0}{B_0^2} \dfrac{dp_0}{dr}$

7) $\sigma(r) = \mu_0 \dfrac{\mathbf{J}_0 \cdot \mathbf{B}_0}{B_0^2}$

equation (5) provides the system

8) $\dfrac{dB_{0\phi}}{dr} = -\sigma B_{0\theta} - g B_{0\phi}$

9) $\dfrac{1}{r}\dfrac{d}{dr}(rB_{0\theta}) = \sigma B_{0\phi} - g B_{0\theta}$

In a real experiment it is unlikely that the pressure gradient could have values significantly greater than zero. Therefore we set $g(r) \leq 0$, so our pressure profiles are monotonic decreasing.

Two important combinations of the equilibrium fields are

10) $F^{m,n} = mB_{0\theta} - n\varepsilon B_{0\phi}$ ;   $\varepsilon(r) = r/R$

11) $G^{m,n} = mB_{0\phi} + n\varepsilon B_{0\theta}$

### 2.2 Perturbations

The *(m, n)* Fourier component of equation (2) is:

12) $\mathbf{J}_0 \times \mathbf{b}^{m,n} + \mathbf{j}^{m,n} \times \mathbf{B}_0 = \nabla p^{m,n}$

Non-linear coupling of the perturbations are neglected, because they would give higher order contributions. Taking the curl of this equation we get rid of the pressure term:

13) $(\mathbf{b}^{m,n} \cdot \nabla)\mathbf{J}_0 - (\mathbf{J}_0 \cdot \nabla)\mathbf{b}^{m,n} + (\mathbf{B}_0 \cdot \nabla)\mathbf{j}^{m,n} - (\mathbf{j}^{m,n} \cdot \nabla)\mathbf{B}_0 = 0$

This equation is coupled to the relations $\nabla \times \mathbf{b}^{m,n} = \mu_0 \mathbf{j}^{m,n}$ and $\nabla \cdot \mathbf{b}^{m,n} = 0$. It is convenient to define the quantity [19]

14) $\psi^{m,n} = -ir b_r^{m,n}$

From the previous equations one gets the harmonics of the poloidal and toroidal field perturbations

15) $b_\theta^{m,n} = \dfrac{1}{m^2 + n^2\varepsilon^2}\left[ -m\dfrac{d\psi^{m,n}}{dr} + n\varepsilon\sigma\psi^{m,n} + n\varepsilon g \dfrac{G^{m,n}}{F^{m,n}}\psi^{m,n} \right]$

16) $b_\phi^{m,n} = \dfrac{1}{m^2 + n^2\varepsilon^2}\left[ n\varepsilon\dfrac{d\psi^{m,n}}{dr} + m\sigma\psi^{m,n} + mg\dfrac{G^{m,n}}{F^{m,n}}\psi^{m,n} \right]$

and the equation which gives the radial profile of $\psi^{m,n}$:

$$17)\ F^{m,n}\left\{\frac{d}{dr}\left[\frac{r}{m^2+n^2\varepsilon^2}\frac{d}{dr}\psi^{m,n}\right]-\psi^{m,n}\left[\frac{1}{r}-\frac{r\sigma^2}{m^2+n^2\varepsilon^2}+\frac{2n\varepsilon(m\sigma+n\varepsilon g)}{(m^2+n^2\varepsilon^2)^2}\right]\right\}=$$

$$\psi^{m,n}\left\{\frac{1}{m^2+n^2\varepsilon^2}\left[rG^{m,n}\frac{d\sigma}{dr}-r\frac{d(gF^{m,n})}{dr}+\frac{2mn\varepsilon G^{m,n}g}{m^2+n^2\varepsilon^2}-rG^{m,n}\sigma g\right]+\frac{2n^2\varepsilon^2 gG^{m,n\,2}}{(m^2+n^2\varepsilon^2)^2 F^{m,n}}\right\}$$

Moreover, writing the relation $\mathbf{B}\cdot\nabla p = 0$ for the *(m, n)* harmonic we have:

$$18)\ p^{m,n} = -\frac{B_0^2 g(r)}{\mu_0 F^{m,n}}\psi^{m,n}$$

All the perturbations are therefore expressed in terms of $\psi^{m,n}$. By defining the quantity

$$19)\ \chi^{m,n} = \left(\frac{r}{m^2+n^2\varepsilon^2}\right)^{1/2}\psi^{m,n}$$

equation (17) is written in a more convenient form:

$$20)\ \frac{d^2\chi^{m,n}}{dr^2}-\chi^{m,n}\left[-\frac{m^4+10m^2n^2\varepsilon^2-3n^4\varepsilon^4}{4r^2(m^2+n^2\varepsilon^2)^2}+\frac{m^2+n^2\varepsilon^2}{r^2}-\sigma^2+\frac{2n\varepsilon(m\sigma+n\varepsilon g)}{r(m^2+n^2\varepsilon^2)}\right]=$$

$$\frac{\chi^{m,n}}{F^{m,n}}\left\{G^{m,n}\frac{d\sigma}{dr}-\frac{d(gF^{m,n})}{dr}+\frac{2mn\varepsilon G^{m,n}g}{r(m^2+n^2\varepsilon^2)}-G^{m,n}\sigma g+\frac{2n^2\varepsilon^2 g}{r(m^2+n^2\varepsilon^2)}\frac{(G^{m,n})^2}{F^{m,n}}\right\}$$

This is the standard force-balance equation for a *(m,n)* radial field in finite-$\beta$ cylindrical RFP. It is possible to demonstrate that this equation is equivalent to the one reported in [20].

### 2.3 Resonant modes

As written in the introduction, a mode *(m, n)* is resonant if the condition $F^{m,n}(r_s^{m,n}) = 0$ is satisfied for $0 < r_s^{m,n} < a$, where *a* is the plasma radius. The force balance equation (17) becomes undefined at the rational surface. Let us discuss the behaviour of the solution near $r_s^{m,n}$. Defining $x = r - r_s^{m,n}$, we take a Taylor expansion of $\sigma$ and *g*:

$$21)\ \sigma(x) = \sigma_0 + \sigma_1 x + o(x^2);\quad g(x) = g_0 + g_1 x + o(x^2)$$

For x → 0 equation (20) reduces to

$$22)\ \frac{d^2}{dx^2}\chi = \chi\left[d + \frac{b}{x} + \frac{c}{x^2}\right]$$

where *d, b, c* are constant (for ease of notation we neglect the superscript *(m,n)*); in particular

$$23)\ b = \frac{1}{(dF/dr)_{r_s}}\left[G\sigma_1 - g_0\frac{dF}{dr} + \frac{2mn\varepsilon G g_0}{r(m^2+n^2\varepsilon^2)} - G\sigma_0 g_0 + \frac{2n^2\varepsilon^2 G^2}{r(m^2+n^2\varepsilon^2)(dF/dr)}g_1 \right.$$

$$\left. - \frac{2n^2\varepsilon^2 G^2 (d^2F/dr^2)}{r(m^2+n^2\varepsilon^2)(dF/dr)^2}g_0\right]_{r_s}$$

$$24)\ c = \frac{1}{(dF/dr)^2}\left[\frac{2n^2\varepsilon^2 G^2}{r(m^2+n^2\varepsilon^2)}g_0\right]_{r_s}$$

The lowerscript $r_s$ means that all the quantities in the right-hand side must be evaluated at the location of the resonant surface. Also, note that $c \leq 0$.

A regular solution near the rational surface is written as [21]:

$$25)\ \chi = |x|^\nu L(x)$$

where

$$26)\ L(x) = L_0 + L_1 x \ln|x| + L_2 x + L_3 x^2 \ln|x| + o(x^2)$$

and $\nu \geq 0$, $L_0 \neq 0$. Inserting in (22) one gets the following conditions

$$27)\ \nu(\nu - 1) = c$$

$$28)\ \nu L_1 = 0$$

$$29)\ L_1(1 + 2\nu) + 2\nu L_2 = L_0 b$$

$$30)\ 2L_3(1 + 2\nu) = L_1 b$$

Condition (27) provides two possible exponents, which are associated to the so-called "large" and "small" solution [21]:

$$31)\ \nu_L = \frac{1-\sqrt{1+4c}}{2}\ ;\quad \nu_S = \frac{1+\sqrt{1+4c}}{2}\ ;\quad \nu_L \leq \nu_S\ ;\quad 0 \leq \nu_{S,L} \leq 1$$

To each of the two exponents is associated a solution, and therefore a set of coefficients $L_0$, $L_1$, …, which will be hereafter labeled with the further lowerscript "$_L$" or "$_S$". We suppose that the condition $1+4c > 0$ holds, otherwise the solution would exhibit oscillating singular behaviour for $x \to 0$. It is possible to show that this condition is just the Suydam criterion:

$$32)\ \frac{rB_\phi^2}{8\mu_0}\left(\frac{q'}{q}\right)^2 > -p'$$

applied at the rational surface. This prescription makes the solution (25) (i.e. the radial field harmonic) a regular function approaching the rational surface.

Nevertheless, we require that also the poloidal, toroidal field and pressure harmonics (15, 16, 18) be regular function for $x \to 0$. By inserting expressions (25, 26) into (15, 16, 18), we note two potentially singular terms; the former arises from the term including the derivative of $\chi$:

$$33)\ \frac{d}{dx}\chi = \nu|x|^{\nu-1}\mathrm{sgn}(x)\cdot L(x) + |x|^\nu\left[L_1 \ln|x| + o(1)\right]$$

The second is due to the $1/F^{m,n}$ term: its contribution is of the form

$$34)\ \frac{g_0}{x}\chi = g_0|x|^{\nu-1}\mathrm{sgn}(x)\cdot L(x)$$

Since these expressions are differently linearly combined in (15, 16, 18), it is not possible that the divergences cancel out each other, and instead they must be regularized separately.

This implies $g_0 = c = 0$, otherwise $0 < \nu < 1$ (see (31)) and the terms with $|x|^{\nu-1}$ would be divergent. The condition $g_0 = c = 0$ imposes that the radial derivative of the zeroth-order pressure vanishes at the rational surface. Moreover, the request of having a monotonic pressure profile brings to the further condition $g_1 = 0$ (there must be a saddle-point, otherwise the rational surface would be the location for a local extreme).

Therefore the two possible exponents of (31) are

35) $v_L = 0$, $v_S = 1$.

The "large" solution gives a singular '$ln|x|$' term in (33). Therefore we set

36) $L_{1,L} = 0$

which, together with (29-30), yields

37) $b = 0$; $L_{3,L} = 0$.

Therefore, the "large" solution reduces to

38) $\chi_L = L_{0,L} + L_{2,L} x + o(x^2)$ .

The condition $b = 0$, together with $g_0 = g_1 = 0$ implies $\sigma_1 = 0$: that is, the radial derivative of the parallel current profile also must vanish at the rational surface.

Note that the "large" solution, which has a <u>finite</u> radial field at the rational surface, is not compatible with the ideal form of the Faraday-Ohm's equation, for which $b_r \propto x \, \xi_r$ near the rational surface [22] ($\xi_r$ is the radial plasma displacement). Our discussion assumes therefore implicitly the effect of resistivity in the Faraday-Ohm's law.

Since we have imposed $b = 0$, for the "small" solution, using again eqns. (28-30), one gets

39) $L_{1,s} = L_{2,s} = L_{3,s} = 0$

which gives

40) $\chi_s = x \left( L_{0,s} + o(x^2) \right)$

In conclusion our requests of regular *(m, n)* harmonics approaching the corresponding rational surface impose the following regularization conditions on the zeroth-order profiles

41) $\left. \dfrac{d\sigma}{dr} \right|_{r_s^{m,n}} = 0$; $\left. \dfrac{dp_0}{dr} \right|_{r_s^{m,n}} = \left. \dfrac{d^2 p_0}{dr^2} \right|_{r_s^{m,n}} = 0$;

The result that the pressure gradient must have a second-order zero at the rational surface was previously obtained in [6] for a general 3-D equilibrium.

With such conditions imposed on the equilibrium profiles the force balance, equation (17), at the mode rational surface degenerates in the identity 0 = 0. It must be solved in the two separate regions between the axis and the rational surface [0, $r_s$[, and between the rational surface and the external boundary ]$r_s$, $r_{shell}$], which is assumed to be given by an ideal shell located at r = $r_{shell}$. The suitable conditions at the origin and at the boundary, and the natural requirement of the continuity of the solution $\psi$ at $r_s$ are imposed. In general it will not be possible to solve the problem with a first radial derivative $d\psi/dr$ continuous at $r_s$. This corresponds to the fact that the coefficient of the small solution may be discontinuous across the rational surface [21].

It is interesting to note that if we retained also quadratic terms in the perturbations in eq. (12), we would add extra singular terms. This has been shown for the force free case [19] (see eqns (A43-A61) in that paper). Anyway these extra singularities can be healed by the same prescription, i.e. flattening $\sigma(r)$.

### 3. A model for $\sigma$ and g

According to the previous discussion, in the presence of saturated resonant modes the quantities $d\sigma/dr$, $g(r)$ and $dg/dr$ must vanish at the mode rational surfaces. In RFX we have $m=0$ modes resonant at the $B\varphi = 0$ surface (the reversal surface), and many $m=1$ modes whose rational surface is inside the reversal surface (internally resonant modes). Therefore we make the *ansatz*

$$42) \quad \frac{d\sigma}{dr}(r) = M(r)f(r), \quad g(r) = M^2(r)h(r)$$

where $M(r)$ is the "regularization" term

$$43) \quad M(r) = q(r)\prod_{n_{min}}^{n_{max}}(1 - nq(r))$$

which automatically satisfies $d\sigma/dr = g(r) = dg/dr = 0$ at the $m=1$ and $m=0$ mode rational surfaces, provided that the shape functions $f(r)$, $h(r)$ (till now undefined) are there regular functions. The regularization prescription (43) is similar to that discussed in [5]. If we want a monotonic decreasing $\sigma(r)$ profile then we set

$$44) \quad M(r) = \left| q(r)\prod_{n_{min}}^{n_{max}}(1 - nq(r)) \right|$$

together with the supplementary constrain $f(r) \leq 0$. Note that, due to the modulus, in this case the second derivative of $\sigma$ is not defined at the rational surfaces. Nevertheless the expressions (21, 22, 23) require the first derivative only. At $r = 0$ for symmetry reasons we have $d\sigma(0)/dr = 0$ and $g(0) = 0$. This implies $f(0) = h(0) = 0$.

Note that $M$ is indeed a function of $q$; as long as $q(r)$ is a monotonic function, we can use $q$ in place of $r$ as independent variable and write

$$45) \quad f(r) = \frac{dq}{dr} w(q), \quad h(r) = \frac{1}{B_0^2} \frac{dq}{dr} u(q)$$

Being $dq(0)/dr = 0$, the symmetry conditions at $r = 0$ are automatically satisfied. From (6, 42, 45) we get

$$46) \quad \frac{d\sigma}{dq} = M(q) w(q) \rightarrow \sigma(q) - \sigma(q_a) = \int_{q_a}^{q} d\bar{q} \, M(\bar{q}) w(\bar{q})$$

$$47) \quad \mu_0 \frac{dp_0}{dq} = M^2(q) u(q) \rightarrow \mu_0 p_0(q) - \mu_0 p_0(q_a) = \int_{q_a}^{q} d\bar{q} \, M^2(\bar{q}) u(\bar{q})$$

where $q_a$ is the value at the plasma boundary, in RFX determined by a graphite wall placed at the radius $r = a$. The graphite wall leans against an inconel vessel located at $r = r_V$, beyond which there is a vacuum region which extends up to the conducting shell placed at $r = r_{shell} = 1.17a$. In a situation of stationary modes helical eddy currents cannot be induced in the vessel, so both $\psi$ and $d\psi/dr$ are continuous at $r = r_V$. There is still the possibility for zeroth-order currents to flow in the inconel vessel; our fluid model fails there, but we are confident that the current amplitude is so small that can be safely neglected. Furthermore zeroth-order currents cannot flow at all in graphite wall [23]. In conclusion we treat the entire region between the plasma boundary and the shell as <u>vacuum</u>:

$$48) \quad \sigma(r) = 0, \quad g(r) = 0 \quad \text{for} \quad a < r < r_{shell}$$

Notice that if we want equation (17) to hold at the boundary $r = a$, then we need the quantities $\sigma$, $d\sigma/dr$, $g(r)$, $dg/dr$ to be defined there. The condition (48) then implies

$$49) \quad \sigma(a) = g(a) = d\sigma(a)/dr = dg(a)/dr = 0$$

From $g(a)= 0= dp_0(a)/dr$ and $p_0(r>a)= 0$ it follows $p_0(q_a)= 0$. Moreover conditions (49) force the following constraints on the functions $w$, $u$:

$$50)\quad w(q_a)=0; \quad u(q_a)=0=\left.\frac{du}{dq}\right|_{q_a}$$

The model depends upon some parameters to be chosen on the basis of supplementary assumptions: in RFX the zeroth-order profiles are characterized by the three experimental parameters

$$51)\quad \Theta =\pi a^2 \frac{B_{0\theta}(a)}{\Phi_t(a)}; \quad F=\pi a^2 \frac{B_{0\phi}(a)}{\Phi_t(a)}; \quad \beta_p = \frac{4\mu_0 \int_0^a p_0(r)r\,dr}{a^2 B_{0\theta}^2(a)}$$

where $\Phi_t$ is the toroidal flux. All the edge magnetic quantities are provided by a direct measure, so we can consider the determination of $\Theta$, $F$ to be exact. Instead, the integral which appears in the expression for $\beta_p$ depends on the assumption made on the pressure profile. At present in RFX $\beta_p$ is computed assuming given polynomial expressions for the electron density and temperature profiles. Our pressure model is different, so we take this "experimental" $\beta_p$ only as a reference value. Note that

$$52)\quad q_a = \frac{a}{R}\frac{F}{\Theta}$$

so $q_a$ can be used in place of F or $\Theta$.

If we want to match a triplet $(F, \Theta, \beta_p)$ of experimental parameters, we need at least three free parameters. Another element to take into consideration is the Suydam criterion (32) for the pressure gradient. In our model this criterion is verified in most of the plasma, due to the prescriptions (42, 43) which flatten the pressure profile on a wide region. A violation could arise in the edge zone, because in our model the pressure gradient is mostly concentrated there. Anyway we have no elements to say that in RFX the Suydam criterion is everywhere fulfilled, and we cannot indeed exclude the presence of localized interchange modes at the very edge of the plasma. This discussion suggests that a fourth free parameter is needed to model the pressure profile near the plasma boundary: this parameter will be tuned in order satisfy, at least marginally, the Suydam criterion there.

According to these arguments and to the condition (50) the simplest model for the functions $w$, $u$ yields the expressions

53) $w(q) = w_0 \left(1 - \dfrac{q}{q_a}\right)^{\xi}$;  $u(q) = u_0 \left(1 - \dfrac{q}{q_a}\right)^{\eta}$;  $\xi > 0$;  $\eta > 1$;

where the four free parameters are $q_a$, the normalization constant $u_0$ and the exponents $\xi$, $\eta$. The normalization constant $w_0$ is not free but depends on ($q_a$, $\xi$) as well as on the definition of the regularization term $M(r)$. This will be demonstrated in the following discussion.

### 4. Getting the solution: numerical scheme
*4.1 Computation of the zeroth-order profiles*

The actual procedure to obtain the zeroth-order profiles for a given triplet *(F, Θ, $\beta_p$)* or *($q_a$, F, $\beta_p$)* is quite involved and in principle, there is not guarantee for the solution to be unique, though we found that all of the admissible solutions are very close between them. First of all, the two possible choices (43, 44) for the regularization term $M(r)$ discriminate between monotonic and not-monotonic $\sigma(r)$ profiles. An integrated analysis [11] of the external magnetic signals and the data provided by a five-chord infrared (FIR) polarimeter indicates that in RFX the $\sigma(r)$ profiles should be <u>very flat</u>, or even <u>hollow</u> with a maximum in the external region of the plasma. In our model the flattening of the $\sigma(r)$ is a direct consequence of the regularization term $M(r)$: the higher the number of modes we have the flatter the profile we get. Moreover with the non-monotonic choice (43) we have a local maximum of $\sigma(r)$ just at the reversal surface where $q(r)=0$. We will investigate both the possibilities (43, 44).

In actual calculations, the product over *(1, n)* modes must-of course-be truncated to a finite number of terms. The number of *m=1* factors in $M(r)$ is determined by the modes experimentally observed. Generally in RFX the dominant internally resonant *m=1* modes have a toroidal number in the range *n=7÷10*. There is also a tail of secondary modes which in some pulses can also extend to high ($n \approx 18,19$) mode numbers. The range [$n_{min}$, $n_{max}$] of factors in $M(r)$ is chosen taking into account also the following two observations:

a) the flattening of the computed $\sigma(r)$, $p_0(r)$ profiles extends beyond the resonance position of the *m=1, n= $n_{max}$* mode. This effect is more pronounced for the pressure, because we have a term $M^2(r)$ there.

b) the zero-gradient constraint is strictly true for the force balance equation (2). As shown in the appendix A, if the small effects of viscosity and inertia are taken into account this constraint slightly relaxes: at the mode rational surface the gradient of the equilibrium current should be not zero but only very small and inversely proportional to the magnetic perturbation amplitude (see formula (A16)). Therefore we decide to include in the product $M(r)$ <u>all</u> of the dominant modes and <u>most</u> of the secondary modes, with the check *a posteriori* that the flat region of $\sigma(r)$ and $p_0(r)$ includes all of the rational surfaces of the modes experimentally observed. The latter condition requires a lower number of factors for $g(r)$ than for $\sigma(r)$, so we leave the possibility that $n_{max}$ appearing in the definition of $M(r)$ could be different for $\sigma(r)$ and $g(r)$. In conclusion, there is some freedom in the definition of $M(r)$, but different plausible choices of the $m=1$ factors do not substantially change the profiles.

Having defined $M(r)$, we set $q_a$ equal to the experimental value (52), and guess a value $\xi>0$ for the exponent of $w(q)$. First of all we compute the $q(r)$ and $\sigma(r)$ profiles: in fact the two equations (8, 9) can be combined to give a single equation for $q(r)$ where the pressure gradient term $g(r)$ does not appear. We combine this equation with the expression for $d\sigma/dr$ to form the system

$$54) \quad \begin{cases} \dfrac{dq}{dr} = \dfrac{2}{r} q \left[ 1 - \dfrac{R\sigma}{2} q \right] - \dfrac{r}{R} \sigma \\ \dfrac{d\sigma}{dr} = w_0 M(r) \dfrac{dq}{dr} \left( 1 - \dfrac{q}{q_a} \right)^\xi \end{cases}$$

The system is solved in the interval $[0, a]$. The apparently more natural direction of integration is from the edge towards the centre, since one knows the starting conditions $q(a) = q_a$, $\sigma(a) = 0$. Note that the condition $\sigma(a) = 0$ forces the normalization constant $w_0$ to be (see (46, 53)):

$$55) \quad w_0 = \dfrac{\sigma(0)}{\int_{q_a}^{q_0} d\bar{q} M(\bar{q})(1 - \bar{q}/q_a)^\xi}$$

that is, the r.h.s. of the latter of (54) depends upon the unknown value of $\sigma$ at the centre.

For this reason it is more convenient to solve eqns. (54) in the direction from the centre towards the edge, using guessed initial values and iterating until the matching at the edge with the boundary conditions is obtained. To this purpose, remember that the relation

$$56) \quad q_0 = \dfrac{2}{R\sigma(0)}$$

must be satisfied. Of course, the value of $\sigma(0)$ must also be consistent with the resonant modes taken in the product (43), so the condition $2(n_{min} - 1)/R < \sigma(0) < 2n_{min}/R$ must hold. We just mention further that the r.h.s of the former of (54), in the limit $r \to 0$, takes the form $0/0$. Thus, some care must be exerted in handling this limit.

The $\sigma(r)$ and $q(r)$ profiles obtained from (54) are then used to determine the zeroth-order magnetic fields (equations (8, 9)). The normalized pressure gradient $g(r)$ is given by

$$57)\ g(r) = u_0 M^2(r) \frac{1}{B_0^2(r)} \frac{dq(r)}{dr} \left(1 - \frac{q(r)}{q_a}\right)^\eta.$$

Remember that the pressure profile is

$$58)\ \mu_0 p_0(r) = u_0 \int_{q_a}^{q(r)} d\bar{q}\, M^2(\bar{q}) \left(1 - \frac{\bar{q}}{q_a}\right)^\eta.$$

In (57) and (58) $q(r)$ is just the solution derived from (54). The parameter $u_0$ is tuned in order to have from the last of (51) a $\beta_p$ close to the "experimental" estimated value. The exponent $\eta$ has little influence on $\beta_p$. It only determines the pressure gradient at the plasma edge and is chosen by comparison with the Suydam criterion.

In general the F and $\Theta$ parameters computed from $B_{0\phi}(r)$, $B_{0\theta}(r)$ are close (they stand in the correct ratio) but not equal to the experimental values. To match the desired values it is sufficient to change slightly the exponent $\xi$ or the number of m=1 resonance (i.e. $n_{max}$) and to repeat the procedure starting from equation (54).

To summarize, our procedure involves solving equations (8, 9, 54) through fitting some free parameters to experimental quantities. Roughly speaking, we can sketch the following recipe:

| Constraints | Related parameter |
|---|---|
| Zero gradient at the mode resonances | Regularization term $M(r)$: $n_{min}$, $n_{max}$ |
| F, $\Theta$ | $q_a$, $\xi$ |
| $\beta_p$. | $u_0$ |
| Suydam criterion | $\eta$ |
| Boundary condition: $\sigma(a)=0$. | $w_0$ |

Table 1. Left column: constraints to be satisfied by the model functions. Right column: terms which are mostly related to the corresponding request of the left column.

We remark that Table I must be considered just as a quick reminder: actually, varying any of the quantity in the right-hand column affects several quantities in left-hand column.

It is convenient to adopt the following normalization for the zeroth-order quantities when working with the above equations:

59) $\hat{r} = r/a; \quad \hat{R} = R/a; \quad \hat{g} = g \cdot a; \quad \hat{\sigma} = \sigma \cdot a;$

60) $\hat{B}_\phi = B_{0\phi} / B_{0\phi}(0); \quad \hat{B}_\theta = B_{0\theta} / B_{0\phi}(0);$

61) $\hat{F}^{m,n} = F^{m,n} / B_{0\phi}(0); \quad \hat{G}^{m,n} = G^{m,n} / B_{0\phi}(0);$

The normalized pressure and the Suydam criterion are written as

62) $\hat{p} = \dfrac{\mu_0 p_0(r)}{B_{0\phi}^2(0)}; \quad -\dfrac{d\hat{p}}{d\hat{r}} < \dfrac{\hat{r}\hat{B}_\theta^2}{8\varepsilon^2}\left(\dfrac{dq}{d\hat{r}}\right)^2;$

From now on, any tilded variable ($\hat{X}$) will implicitly refer to a normalized quantity.

*4.2 Perturbation quantities*

The radial profile of the various *(m, n)* harmonics is obtained from eq. (20), which is solved using the zeroth-order profiles for σ*(r)* and *g(r)*, computed according to the method just outlined. The edge boundary conditions, which determine the perturbation amplitudes, are given by the experimental measurements, in RFX available at the shell radius $\hat{r}_{shell} = r_{shell}/a = 1.17$:

63) $\chi^{m,n}(\hat{r}_{shell}) = 0; \quad \dfrac{d\chi^{m,n}(\hat{r}_{shell})}{d\hat{r}} = \dfrac{\hat{r}_{shell}^{1/2}\left(m^2 + n^2\varepsilon_{shell}^2\right)^{1/2}}{n\varepsilon_{shell}} a^{3/2} b_\phi^{m,n}(\hat{r}_{shell});$

The first condition (63) is strictly true only for an ideal shell. The RFX shell is thick but not exactly ideal, since a slow penetration of the radial field is observed. This is shown by the measurements of two poloidal arrays of radial field pick-up coils placed on the inner surface of the shell. At present toroidal arrays of radial field probes, which are necessary to obtain the harmonics $b_r^{m,n}$ (proportional to $\chi^{m,n}$) at a given radius, are not available. Therefore the simplest way to set the initial value for $\chi^{m,n}$ is to adopt the ideal shell approximation (assumed in all of the following examples).

The second condition (63) is derived by a combination of (16) and (19) applied in vacuum, and by the ideal shell constraint $\chi^{m,n}(r_{shell})=0$. The harmonic $b_\phi^{m,n}$ is indeed <u>measured</u> at the shell (for the *m= 0, 1* modes) by two toroidal arrays of 72 equally spaced toroidal field pick-up coils placed at opposite poloidal angles [24].

The condition near the origin is $\chi^{m,n}(\hat{r}) \propto \hat{r}^{|m|+1/2}$ for $m \neq 0$ and $\chi^{0,n}(\hat{r}) \propto \hat{r}^{3/2}$ for $m=0$. Equation (20) is solved in two distinct regions: between the origin and the mode rational surface radius $r_s^{m,n}$, and between $r_s^{m,n}$ and the shell. The matching of the two solutions is obtained imposing the continuity of $\chi^{m,n}$ at $r_s^{m,n}$. In general the first radial derivative $d\chi^{m,n}/dr$ is discontinuous there.

The phase $\varphi^{m,n}$ of the harmonic $\psi^{m,n}$ is taken constant with respect to $r$, so we can write

64) $\psi^{m,n}(r,t) = \tilde{\psi}^{m,n}(r,t) e^{i\varphi^{m,n}(t)}$.

Note that the mode amplitude $\tilde{\psi}^{m,n}$ is a real solution of equation (17). The discontinuity of the first radial derivative of $\psi$ across the rational surface is quantified by the real parameter

65) $E^{m,n} = \dfrac{\hat{r}_s^{m,n}}{\tilde{\psi}^{m,n}(\hat{r}_s^{m,n})} \dfrac{d\tilde{\psi}^{m,n}}{d\hat{r}} \Bigg|_{\hat{r}_s^{m,n}-}^{\hat{r}_s^{m,n}+}$

By integrating Ampere's law

66) $\mu_0 j_\theta^{m,n} = -i\dfrac{n}{R} b_r^{m,n} - \dfrac{\partial b_\phi^{m,n}}{\partial r}$; $\quad \mu_0 j_\phi^{m,n} = \dfrac{1}{r}\dfrac{\partial(r b_\theta^{m,n})}{\partial r} - i\dfrac{m}{r} b_r^{m,n}$ ;

in a narrow region $[r_s^{m,n} - \delta, r_s^{m,n} + \delta]$ around the mode rational surface, we can get the poloidal and toroidal components of the helical current sheet which flows there:

67) $\mu_0 \displaystyle\int_{r_s^{m,n}-\delta}^{r_s^{m,n}+\delta} j_\theta^{m,n} dr = -b_\phi^{m,n}\Big|_{r_s^{m,n}-\delta}^{r_s^{m,n}+\delta} + o(\delta) = -\dfrac{n\varepsilon_{rs}}{a(m^2 + n^2\varepsilon_{rs}^2)} \dfrac{\tilde{\psi}(\hat{r}_s^{m,n})}{\hat{r}_s^{m,n}} E^{m,n} + o(\delta)$

68) $j_\phi^{m,n} = \dfrac{m}{n\varepsilon} j_\theta^{m,n}$

In the last equality of (67) we have taken into account that the third term in the square bracket of (15, 16) is zero at the rational surface, because $g(r)$ has a second order zero there.

### 5. Examples of application of the method
*5.1 Zeroth-order profiles: monotonic $\sigma(r)$*

Table 2 reports some examples of zeroth-order profiles computations with the assumption of a monotonic $\sigma(r)$, therefore using the definition (44) for $M(r)$. Note that in this case $w_0 > 0$, so $d\sigma(r)/dr$ is negative everywhere, being $dq/dr<0$.

| Shot | F | $\Theta$ | $\beta p$ | $(1, n)\sigma$ | $(1, n)$ g | $\hat{\sigma}(0)$ | $\xi$ | $\hat{u}_0$ | $\eta$ |
|---|---|---|---|---|---|---|---|---|---|
| 11029 t=30 | -0.119 | 1.408 | 0.060 | 7-14 | 7-12 | 2.83 | 0.4 | 2200 | 1.3 |
| 11029 t=35 **PPCD** | -0.224 | 1.476 | 0.056 | 7-10 | 7-9 | 2.844 | 0.05 | 330 | 1.2 |
| 8763 t=66 | -0.163 | 1.416 | 0.060 | 7-13 | 7-13 | 2.7876 | 0.35 | 1100 | 1.8 |
| 8763 t=73 **α-mode** | -0.217 | 1.472 | 0.052 | 7-10 | 7-10 | 2.854 | 0.06 | 450 | 1.6 |
| 8073 t=31.5 | -0.251 | 1.49 | 0.090 | 7-14 | 7-14 | 2.792 | 0.6 | 250 | 2.2 |

Table 2. Plasma parameters as guessed from our model for a set of experimental conditions. F, $\Theta$, and poloidal beta are from experiment and must be matched by the model. The other columns contain, from left to right: the modes retained in the computation of $\sigma$ and of pressure, the fitting parameters appearing in Eqns (54-57).

The fifth and sixth columns contain the range of $m=1$ factors used in the $M(r)$ terms for $\sigma(r)$ and $g(r)$ respectively. The $m=0$ factor is always present.

All the calculations have been performed with the "*Mathematica 4.0*" software. Note that the exponents $\xi$, $\eta$ do not vary too much from shot to shot. In the first column the time is given in *ms*.

Our model provides profiles which fit the global/edge measured parameters $(F, \Theta, \beta_p)$. Comparing our results with whole radial shape of the profiles is at present impossible, since in RFX we have not a shot-by shot measurement of these profiles. Anyway, as told in sec 4.1, the first experimental reconstruction of $\sigma(r)$ indicates a flat profile (see fig. 4 of Ref. [11]), which is largely compatible with the plots shown below. The same is true for the pressure, even if only the electronic component is measured (see fig. 4 of Ref. [14]).

Pulse 11029 at t=30ms

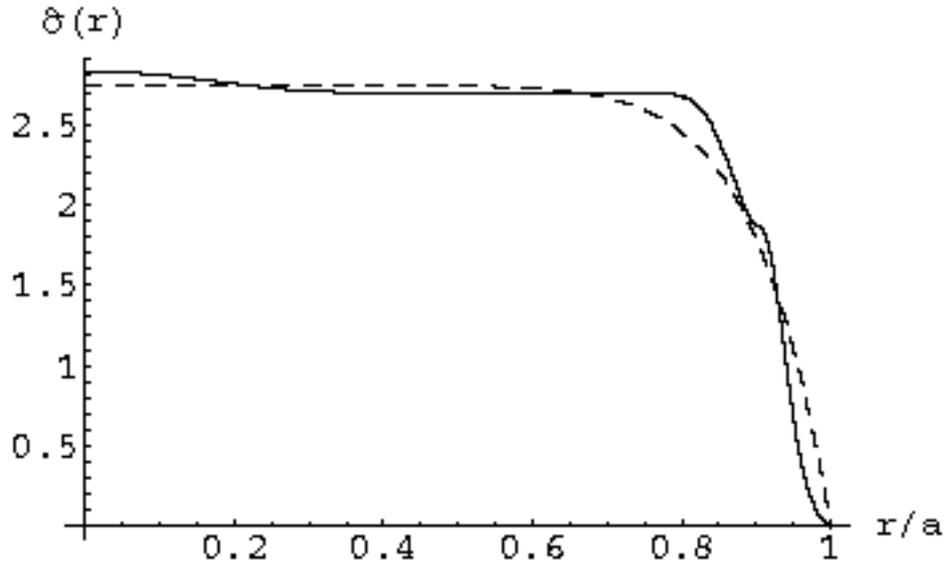

Fig. 1. The normalized σ profile. Pulse 11029 at t=30ms (standard conditions). Continuous line present model; dashed-line $\alpha$-$\Theta_0$ model

In Fig. 1 the continuous line is the result of our model while the dashed line is the reconstruction provided by standard $\alpha$-$\Theta_0$ model. The differences between the two models arise in the central part of the plasma, where our model predicts a weak increase of $\sigma(r)$ towards zero, and in the external region where we obtain a more extended flat zone, a further flattening in correspondence of the *m=0* resonance (*r≈ 0.9a*) and a smoother behaviour near the plasma boundary. The flat region *0.3 < r < 0.8* corresponds to the location of *m=1* modes. Figs. (2-4) show the other fundamental quantities calculated within the model.

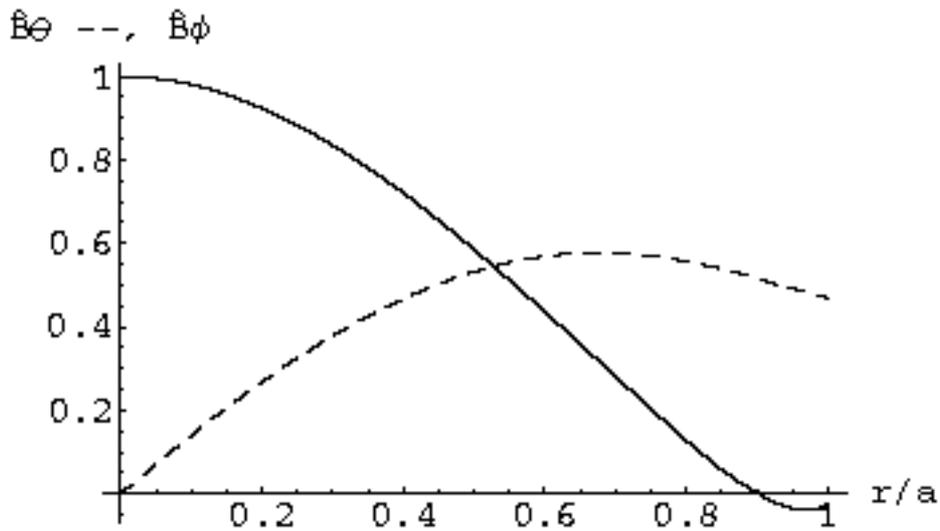

Fig. 2: The normalized toroidal and poloidal field profiles.

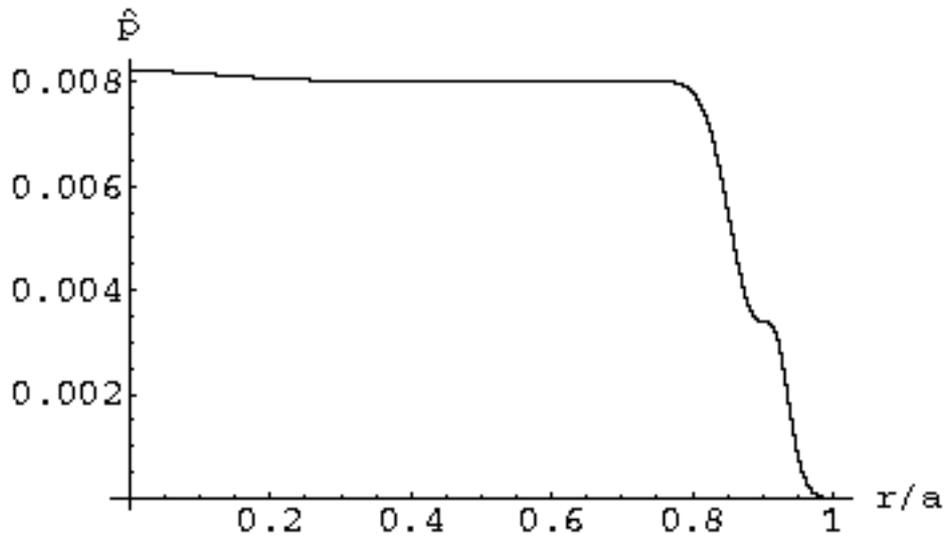

Fig. 3: The normalized pressure profile. The flattening due to the *m=0* modes at *r≈ 0.9a* is clearly evident

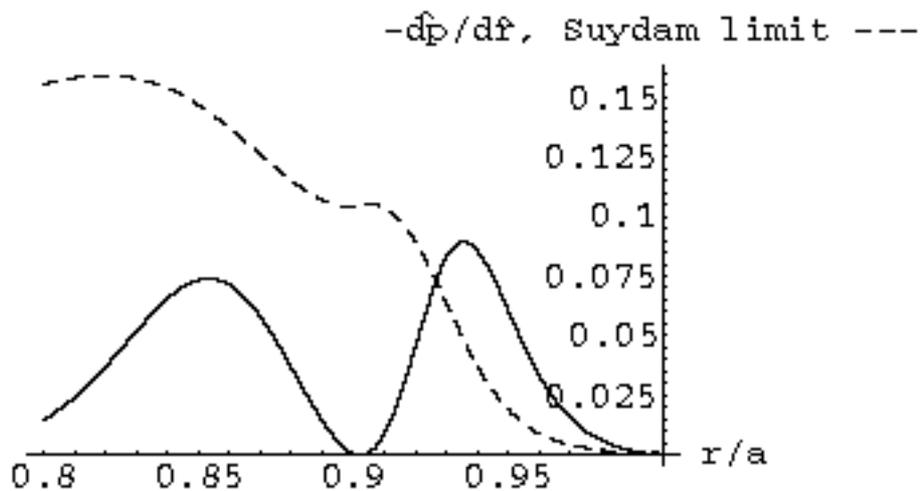

Fig. 4. Here the comparison, in the external region of the plasma, between the normalized pressure gradient (continuos line) and the Suydam limit (dashed line) is shown: the violation of this limit is at the very edge. The agreement could be improved by increasing the exponent η.

Pulse 11029, t = 35 ms, during PPCD

The same shot is considered when the PPCD [13] is active. During the PPCD some of the secondary *m=1* modes disappear. This entails a reduction of $n_{max}$ in *M(r)* and a consequent steepening of the σ*(r)* profile (fig. 5).

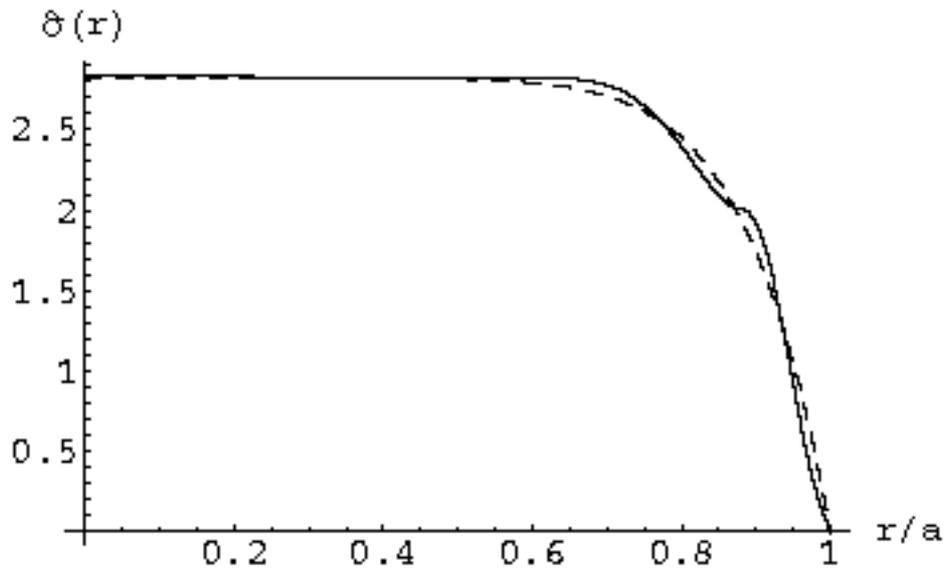

Fig. 5. The normalized σ profile. Pulse 11029 during PPCD. Continuous line present model; dashed-line α-Θ$_0$ model

The peaking of the profile is also predicted by the α-Θ$_0$ model. A similar peaking is apparent in the pressure profile, and it is also observed experimentally [14]. As discussed is the introduction this confirms the existence of a connection between the number of the instabilities and the shape of the zeroth-order profiles. During the PPCD the activity of the dynamo modes is weaker and consequently the profile is less relaxed (steeper).

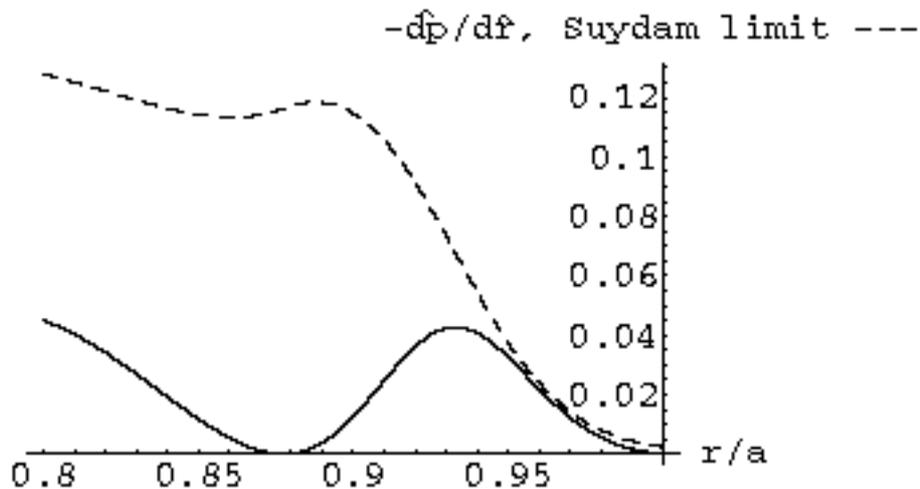

Fig. 6. During the PPCD our model satisfies the Suydam criterion also at the very edge of the plasma.

Alike the PPCD, the so called α-mode is characterized by a decrease of the secondary m=1 modes and by a steepening of the profiles. In the α-mode reported in the table our model satisfies the Suydam criterion also at the very edge of the plasma.

*5.2 Zeroth-order profiles: non-monotonic σ(r)*

As mentioned before in sec.4.1, there are evidences supporting the possibility for σ profiles to be non-monotonic in RFX plasmas. In our formalism, it means using Eq. (43) instead of (44) in computations. Note that in this case $w_0<0$, because of the significant negative contribution of $M(r)$ to the integral (55) in the region between the reversal and the edge. With this kind of profile we found that the convergence of equations (8, 9, 54) to the experimental desired values of $(F, \Theta, \beta_p)$ is not guaranteed. We relate this to the following physical aspects of the problem:

(I) we have found that a necessary condition is a decreasing σ(r) between the origin and the first $m=1$ rational surface. With the monotonic profile this condition is automatically fulfilled. Instead, using the definition (43) it is satisfied only with an odd number of $m=1$ mode. This suggest that the non-monotonic model for σ(r) should be refined including an extra factor, e.g.:

$$69)\quad \frac{d\sigma}{dr} = w_0 q(r) \prod_{n_{\min}}^{n_{\max}} (1 - nq(r)) \frac{dq}{dr} \left(1 - \frac{q}{q_a}\right)^{\xi} \cdot \left[\mathrm{sgn}(1/n_{\min} - q)\right]^{n_{\max} - n_{\min}}$$

(II) There should not be too much distance between the rational surface corresponding to $n_{max}$ and the reversal position, otherwise σ(r), which has a positive derivative in this region, could grow to unreasonable values at the reversal. This requires a configuration with a considerable number of secondary $m=1$ modes, in other word a sufficiently relaxed profile.

(III) Since in the non-monotonic case the maximum of σ(r) is reached at the reversal surface and it is unlikely that too much current be driven at the very edge of the plasma, the reversal should be at a suitable distance from the plasma boundary, i.e. the F parameter should be sufficiently negative, to make the profile plausible.

Table 3 lists some specific cases. From fig.7 one can notice a significant departure from the $\alpha$-$\Theta_0$ reconstruction.

| Shot | F | Θ | βp | (1,n)σ | (1,n)g | σ̂(0) | ξ | û₀ | η |
|---|---|---|---|---|---|---|---|---|---|
| 8073 t=31.5 | -0.251 | 1.49 | 0.09 | 7-15 | 7-13 | 2.832 | 0.96 | 550 | 2.4 |
| 8069 t=40 | -0.391 | 1.58 | 0.08 | 7-13 | 7-11 | 2.875 | 1. | 120 | 2 |
| 14170 t=40 | -0.249 | 1.463 | 0.06 | 7-14 | 7-12 | 2.838 | 1. | 350 | 2 |

Table 3. Same symbols of Table 2 but for the non-monotonic current profile.

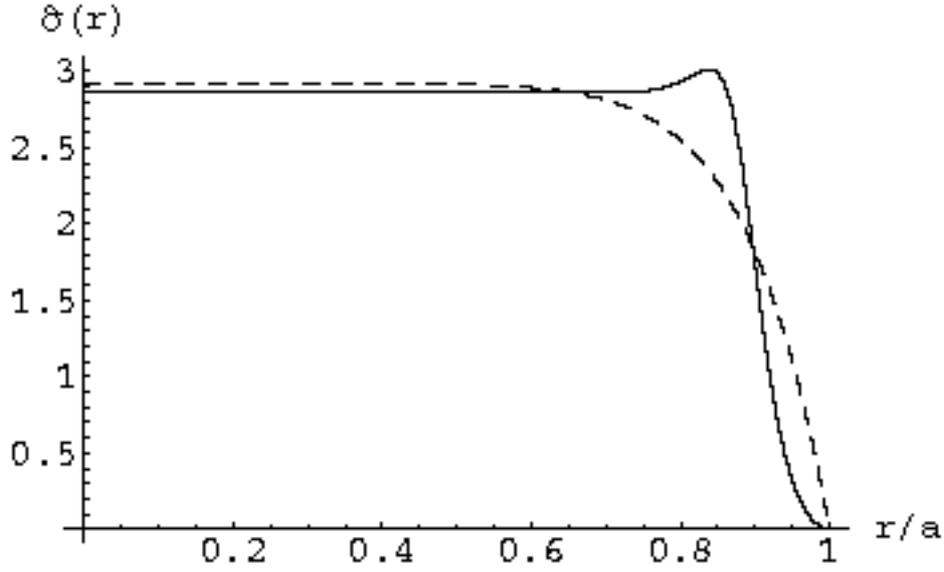

Fig. 7: Shot 8069, t=40ms. Continuous line present model; dashed-line α-Θ$_0$ model.

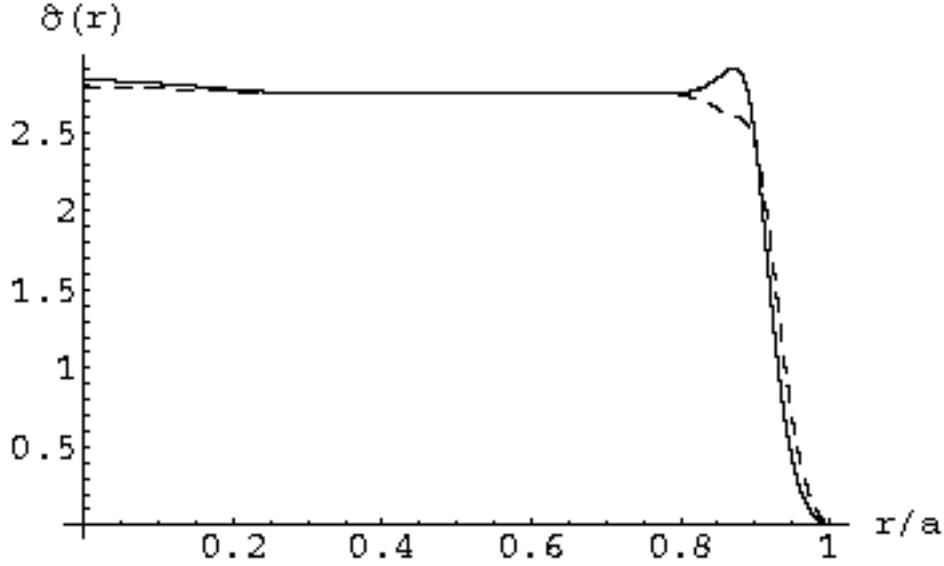

Fig. 8: Shot 8073, t=31.55ms The two kind of profiles admissible in our model, monotonic (dashed line) and not-monotonic (continuous line), are compared. In this case there is not a substantial difference apart in correspondence of the *m=0* mode.

*5.3 Examples of perturbation profiles*

Let's examine the radial profile of the perturbation associated to the typical dominant *m=1, n=8* mode. Figures 9, 10, 11, feature respectively the amplitudes $\tilde{\psi}^{1,8}(r)$, $\tilde{b}_r^{1,8}(r)$, $\tilde{b}_\phi^{1,8}(r)$ in the pulse 11029 at t=30ms. The zeroth-order profile has already been determined (first row of Table 2). It is indeed a huge amplitude mode, with an edge measured value $b_\phi^{1,8}(1.17)= -6.2mT$.

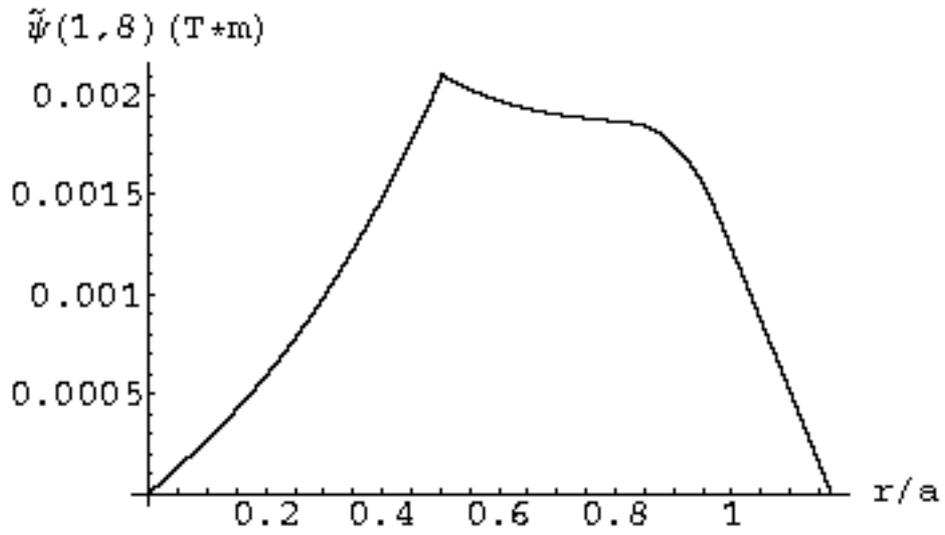

Fig. 9: Shot 11029 t=30ms. The units are (T·m)

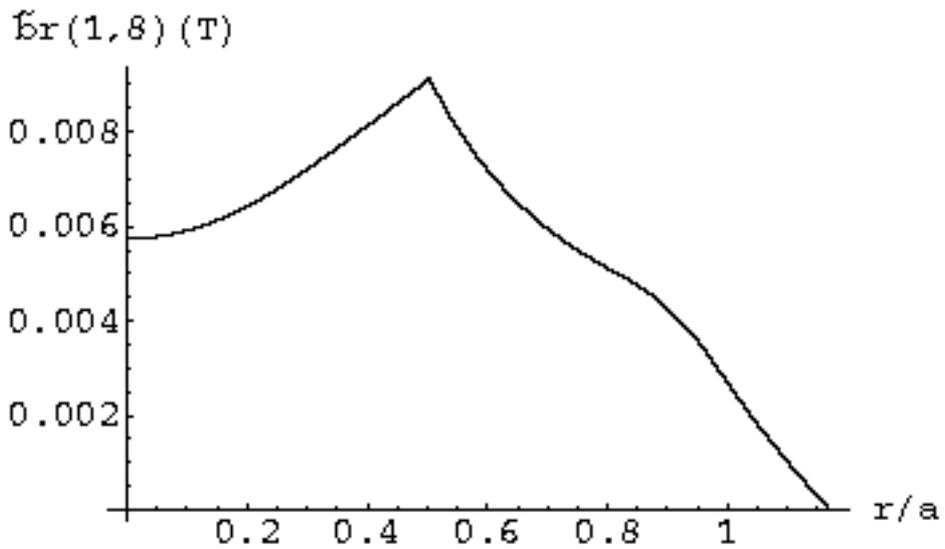

Fig. 10: The radial field amplitude profile for the same shot of fig.9

Note the discontinuity of the first derivative across the mode rational surface. We have found that $E^{1,8}=-1.98$. The radially integrated poloidal and toroidal component of the current sheet are therefore $J_\theta^{1,8}= 7165\ A/m$, $J_\phi^{1,8}=7781 A/m$

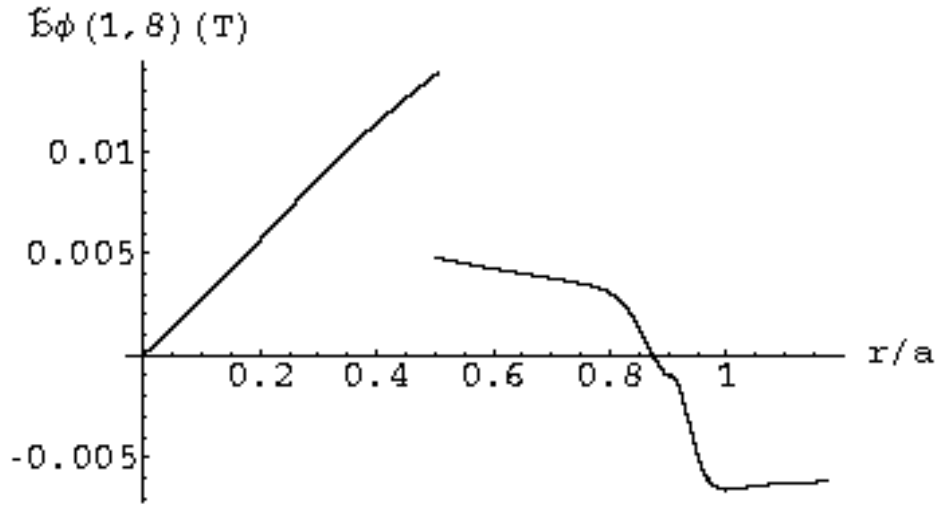

Fig. 11: The toroidal field amplitude profile for the mode (1,8). The jump at the rational surface is a consequence of the presence of the $d\psi/dx$ discontinuity there (i.e., a current sheet flows at the rational surface).

For the same shot and time we consider the mode $m=1$, $n=14$. In this case the amplitude is lower: $b_\phi^{1,14}(1.17)= -1.75 \ 10^{-3}$ T. The corresponding profiles are shown below.

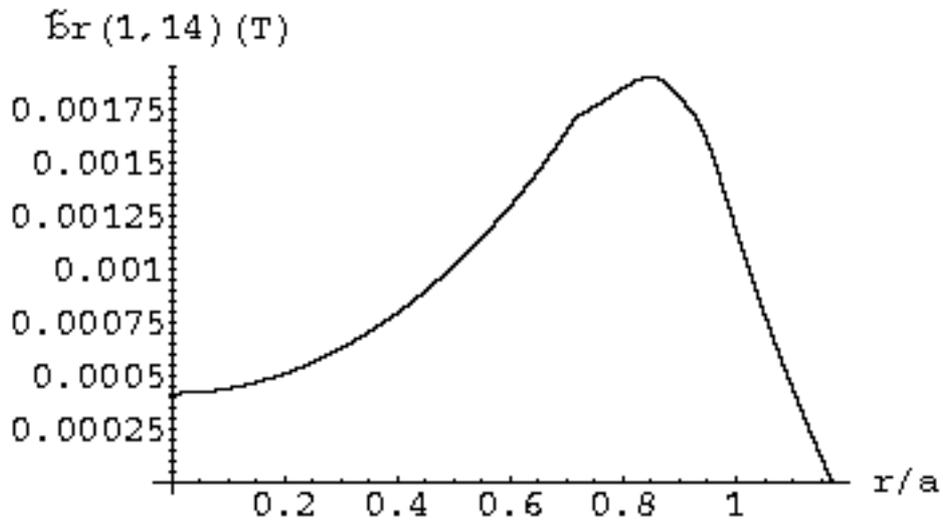

Fig. 12. Radial perturbation for the m=1, n=14 mode.

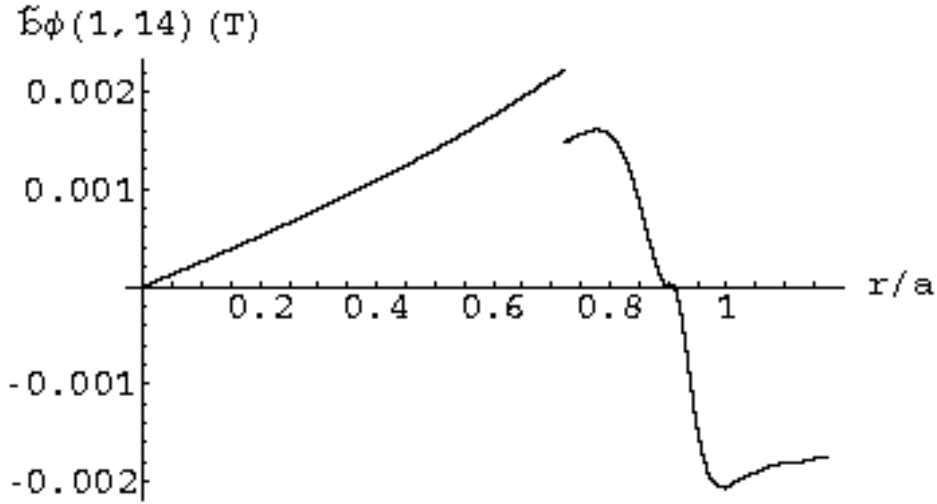

Fig. 13. Toroidal perturbation for the m=1, n=14 mode.

We find $E^{1,14}= -0.78$; $J_\theta^{1,14} = 391$ A/m, $J_\phi^{1,14} = 169$ A/m.

At present the ideal shell constraint can not be overcome, because the lacking of an adequate set (i.e. toroidal arrays) of radial field probes prevents the direct measure of $\chi^{m,n}$ there. Anyway a slight improvement of this condition can be realized making use of two poloidal arrays of 16 radial field pick-up coils. In this way it is possible to estimate the penetration of the shell by the perturbation. Removing the ideal shell condition we find small variations only for the modes which resonate in the outer region of the plasma. For the last example *m=1, n=14* we find a small increase of the perturbation value at the rational surface and a positive value $E^{1,14}= 0.236$.

Note that in all these examples we have found $E^{m,n} \neq 0$, so the jump of $d\psi/dx$ across the rational surface is finite. In particular we get $E^{m,n} < 0$ for the dominant *m= 1* modes and $E^{m,n} > 0$ for the *m= 0* modes (see fig.14). This is in agreement with a previous analysis on RFX profiles [25]. The parameter $E^{m,n}$ is related to the stability of the mode [3, 21]. Note that the definition (65) is suitable for a finite-β plasma if the pressure gradient vanishes at the rational surface, as imposed in our equilibrium model. The evolution of an island associated to a tearing mode island is described by the Rutherford theory [26]. It this case marginal stability should be equivalent to the condition $E^{m,n} \approx 0$. A more rigorous version of this theory [27] indicates that a mode with $E^{m,n} >0$ grows until its amplitude reaches a saturation value which depends on $E^{m,n}$. The fact that we get $E^{1,n} <0$ does not seem in agreement with these predictions. We do not think to solve this apparent contradiction here. Anyway let us point out two observations: 1) The magnetic field in the plasma core of a RFP is stochastic, due to the effect of overlapping magnetic islands of the *m= 1* modes. Probably the Rutherford-type single-helicity theories should not be applied to the stochastic region of the *m=1* modes. 2) If our model is correct, the growth of many instabilities entails an important modification of the zeroth-order profiles. When a condition of marginal stability is finally reached the final relaxed profiles could be quite different from the ones which have given origin to the modes. Therefore also the parameters $E^{m,n}$ should change in time according to the zeroth-order profiles modifications: the value $E^{1,n} < 0$ computed for the steady state should not mean that the $m=1$ modes are stable; instead it is a consequence of the flattening of the zeroth-order profiles in the central region of the plasma. Take into account that our zeroth-order profiles are close to the physically realized ones in RFX, and that any plausible equilibrium profile for RFX would give $E^{m,n} < 0$ for the dominant $m=1$ modes.

### 5.4 $m=0$ magnetic island in RFX

As further example we present the reconstruction of the island associated to the $m=0$ modes. The RFX discharges are characterized by an important $m=0$ perturbation, which contains many toroidal Fourier components locked in phase together [16, 18]. For sake of simplicity the effect due to the $m=1$ perturbations is not taken into account. In the presence of $m=0$ perturbations only, the poloidal co-ordinate can be ignored and the magnetic field can be expressed in terms of the toroidal-flux function:

$$70)\ \mathbf{B} = \frac{1}{2\pi r}\nabla\Psi_T \times \mathbf{e}_\theta + B_\theta\,\mathbf{e}_\theta$$

From (70) we get:

$$71)\ b_r = -\frac{1}{2\pi rR}\frac{\partial\Psi_T}{\partial\phi};\quad B_\phi = \frac{1}{2\pi r}\frac{\partial\Psi_T}{\partial r}$$

From (14, 16) the total radial and toroidal $m=0$ perturbations are written:

$$72)\ b_r^{m=0}(r,\phi) = \frac{2}{r}\sum_{k>0}\tilde{\psi}^{0,k}(r)\sin\!\left[k\phi - \varphi^{0,k}\right]$$

$$73)\ b_\phi^{m=0}(r,\phi) = 2\sum_{k>0}\frac{1}{k\varepsilon}\frac{d\tilde{\psi}^{0,k}}{dr}\cos\!\left[k\phi - \varphi^{0,k}\right]$$

A factor 2 appears in front of the summations because we have to include both the *(0, k)* and the complex conjugate *(0, -k)* modes. Writing $B_\phi(r,\phi) = B_{\phi 0}(r) + b_\phi^{m=0}(r,\phi)$ the toroidal flux is obtained from (71, 72, 73):

$$74)\ \Psi_T(r,\phi) = 2\pi \int_0^r \rho\, B_{0\phi}(\rho)\,d\rho + 4\pi R \sum_{k>0} \frac{\tilde{\psi}^{0,k}(r)}{k} \cos(k\phi - \varphi^{0,k})$$

The shot taken into consideration is the 8071, at $t = 9$ ms. The zeroth-order profiles are obtained with our model. The reversal surface is predicted at $r_* = 0.878 \cdot a$. The toroidal profile of the perturbation (the toroidal field component) measured at the shell, is given in Fig. 14.

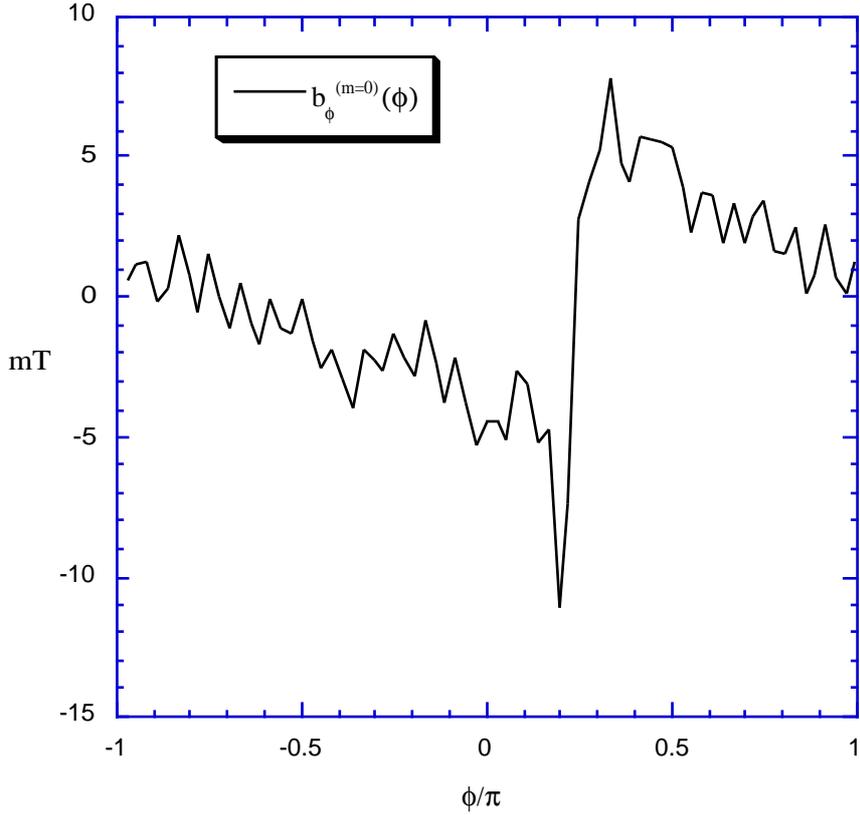

Fig.14: m=0 toroidal field perturbation $\phi$-profile measured at the shell radius

The harmonics amplitude of this profile provide the edge boundary conditions for the perturbation equation (eq. (20)). The computed radial profile of $\tilde{\psi}^{0,k}(r)$ is shown in Fig.15 for $k=9$. Note that the value $\tilde{\psi}^{0,k}(r_*)$ at the reversal surface is much smaller than the maximum which is located in the very edge region of the plasma. This is related to the presence of a vacuum gap between the plasma and the shell, which covers the region *1< r/a <1.17*. In fact if the shell were just at the plasma edge *r/a= 1* we would obtain a profile more similar to those of the secondary *m=1* modes.

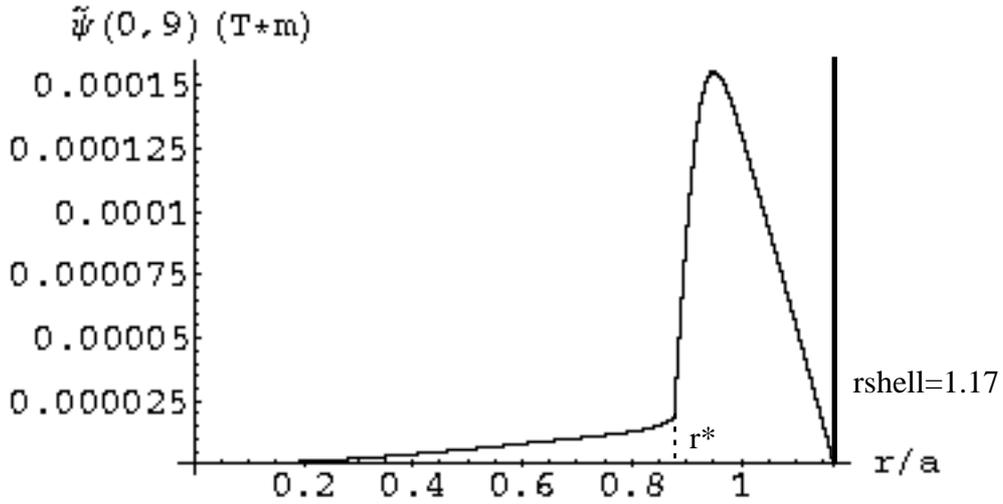

Fig. 15: Radial profile of the perturbation amplitude ψ for the m=0 n=9 mode.

Moreover from the same Fourier decomposition of the signal displayed in fig,14 we can get the modes phases $\varphi^{0,k}$, to be inserted, together with the amplitudes, in eq. (74). Note that the steep gradient there appearing is due to the phase-locking of the modes:

75) $\varphi^{0,k} = k \cdot \phi_0 - \Delta_0; \quad \Delta_0 \approx \pi/2;$

The locking angle $\phi_0$ coincides with the position in which the steep gradient crosses zero. The parameter $\Delta_0$ has a value about $\pi/2$ which corresponds to the minimum non-linearly generated torque at the reversal surface [16].

The contour plot of the flux function defined by eq. (74) (fig.16) displays a toroidal cross-section of the flux-surfaces located near the reversal surface. The harmonics with *k=1,…10* are taken into account. It is apparent the existence of an island, highly distorted in the toroidal direction. The island width is about *0.15× a=6.86cm*. Moreover the structure is not symmetric around the reversal surface due to the fact that the *m=0* perturbation is mostly concentrated in the very edge of the plasma.

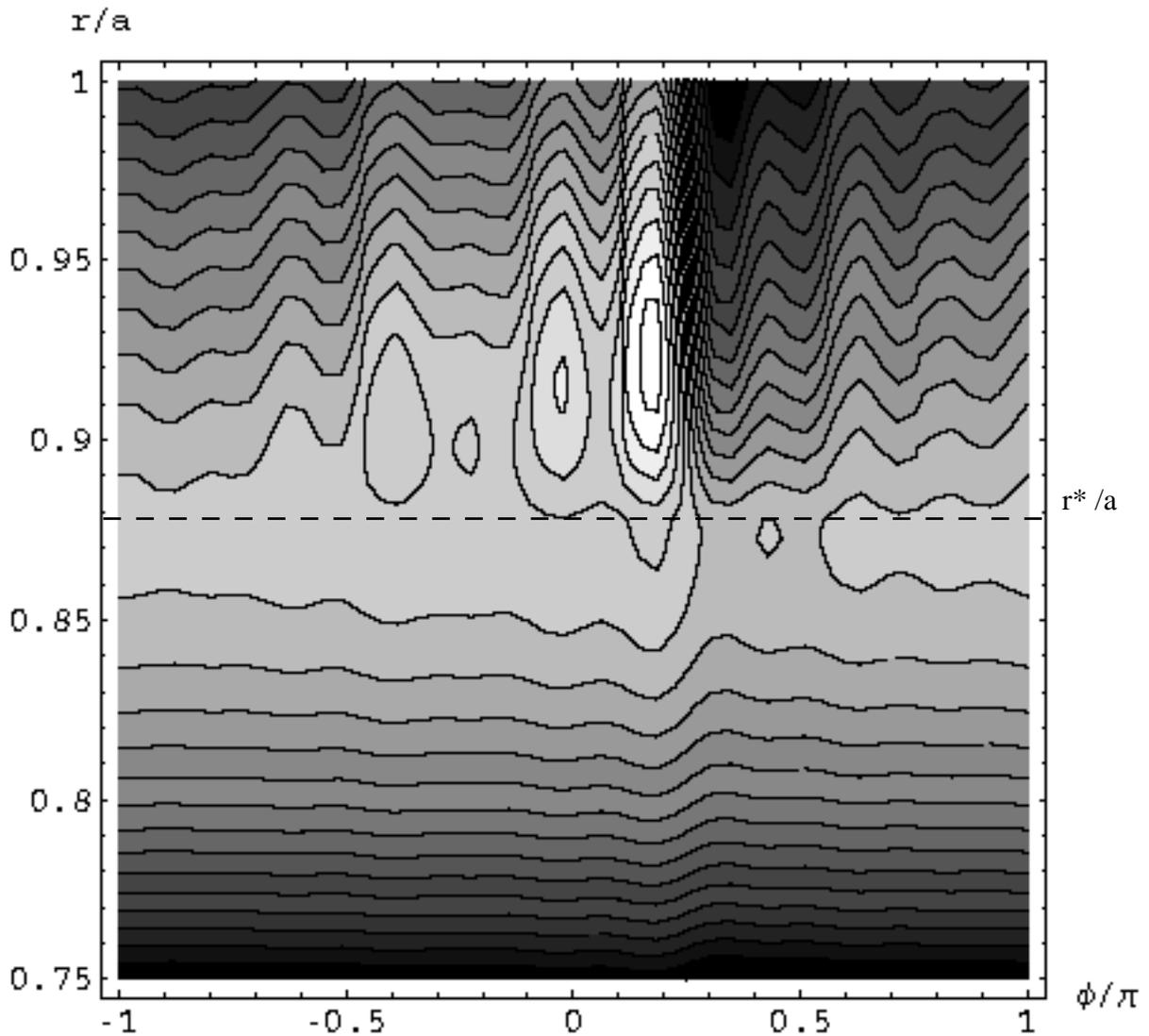

Fig.16: Toroidal cross section of the flux surfaces near the reversal surface, using the computed radial profile $\tilde{\psi}^{0,k}(r)$. The contributions for the k=1,..,10 harmonics have been summed up.

## 6. Conclusions

The equilibrium in a finite-$\beta$ RFP plasma in the presence of saturated-amplitude stationary tearing modes has been investigated analytically and numerically. The singularities of the force balance equation at the modes rational surfaces have been dealt with by a proper regularization of the zeroth-order cylindrically symmetric profiles: the gradients of the pressure and parallel current density ($\sigma$) are forced to be equal to zero there. This rule has been combined with some considerations about general bounds plasmas are constrained to, to guess equilibrium profiles which match the global discharge parameters *(F, $\Theta$, $\beta_p$)*. The combined effects of a large number of modes makes the $\sigma$ and pressure profiles flat over a large portion of the plasma, with a steep gradient in the edge region, although non-monotonic $\sigma$ profiles, with a maximum at the reversal surface, are also allowed in principle. A corollary of our regularization rule is that to obtain steeper profiles an RFP needs to reduce drastically the dynamo modes amplitude. This is not surprising, if we think that all the improved regime of operations in RFX ($\alpha$-mode [12], PPCD [13], "single-helicity" states [28]) are characterized by a reduced mode amplitude.

At present, in RFX we have not a shot-by shot measurement of $\sigma$ and pressure profile. Anyway, the first experimental analysis of $\sigma(r)$ [11] is largely compatible with our model. The same is true for the pressure, even if only the electronic component is measured [14]. When some of the modes are suppressed, the profiles both in our model and in experiment become steeper.

The model can be regarded as a method to obtain the final stage of relaxation process produced by the tearing modes. This equilibrium model allows the use of a relatively simple equation to get the various perturbations profiles. In this way a complete magnetic reconstruction is obtained. Some interesting features, for example the structure of the *m=0* island have been already obtained. The possible implications of the magnetic reconstruction on other aspects, such as the transport, can be investigated.

Some of the arguments which have led us to the results shown in previous sections, are forcefully semi-quantitative, therefore some room to arbitrariness is left. In particular, we think that finding further constraints allowing to better define "shape" functions *w, u* (53), would be a possible improvement. Furthermore, the present version of our model allows only an "on/off" picture of the flattening effect of the resonant modes over the equilibrium. In a more sophisticated model a correlation between the mode amplitude and the width of the induced flattening should be self-consistently taken into account.

## Appendix A

The inclusion of small effects like inertia and viscosity in the equilibrium motion equation

A1) $\quad \rho \mathbf{V} \cdot \nabla \mathbf{V} = \mathbf{J} \times \mathbf{B} - \nabla p + \nu \nabla^2 \mathbf{V}$

does not substantially change our constraint of zero-gradient for the zeroth-order quantities at the rational surfaces. In this case the system of equations (A1, 3) is closed adding the steady state (for stationary modes) Faraday-Ohms's law

A2) $\quad \nabla \times (\mathbf{V} \times \mathbf{B}) + \dfrac{\eta}{\mu_0} \nabla^2 \mathbf{B} = 0$

where density $\rho$, viscosity $\nu$ and resistivity $\eta$ are taken as constant. Assuming a slab geometry and simple field profiles, the equations become analytically manageable. Despite the simplification all the key features of the problem are still present.

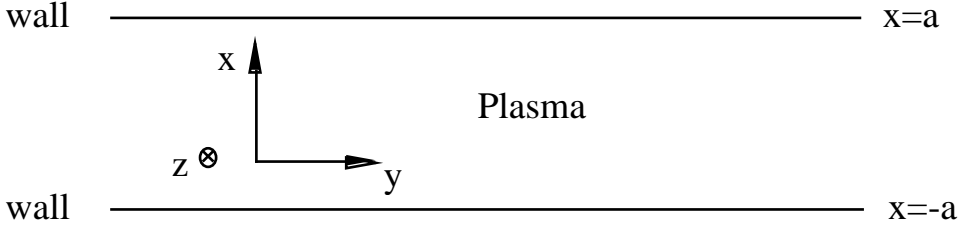

Fig. A1: sketch of the simplified geometry used in the calculation

Let's set $\partial/\partial z = 0$ for both the zeroth-order and the perturbed quantities. We chose the following zeroth-order fields:

A3) $\mathbf{B}_0 = [0, B_y(x), 0]$ ; $\mathbf{V}_0 = [0, V_y(x), 0]$.

Therefore

A4) $\mu_0 \mathbf{J}_0 = [0, 0, B'_y] = \mu_0 J_{0z} \mathbf{z}$ ; $\mathbf{W}_0 = \nabla \times \mathbf{V}_0 = [0, 0, V'_y]$

The perturbation is written $f(x,y) = f_k(x) e^{iky}$. The physical perturbation is the real part of this expression. In the following we will refer to the k-th harmonic of the perturbed quantities. The perturbed velocity and magnetic field are assumed to have the simplest possible divergence-free form:

A5) $\mathbf{b} = \nabla \psi \times \mathbf{z} = [ik\psi, -\psi', 0]$ ; $\mathbf{v} = \nabla \phi \times \mathbf{z} = [ik\phi, -\phi', 0]$

Therefore

A6) $\mu_0 \mathbf{j} = -\nabla^2 \psi \, \mathbf{z}$ ; $\mathbf{w} = \nabla \times \mathbf{v} = -\nabla^2 \phi \, \mathbf{z}$ ; $\nabla^2 \equiv \dfrac{d^2}{dx^2} - k^2$

The curl of the first-order component of (A1) is:

A7) $\rho \left[ (\mathbf{V}_0 \cdot \nabla)\mathbf{w} + (\mathbf{v} \cdot \nabla)\mathbf{W}_0 - (\mathbf{W}_0 \cdot \nabla)\mathbf{v} - (\mathbf{w} \cdot \nabla)\mathbf{V}_0 \right] =$

$= (\mathbf{b} \cdot \nabla)\mathbf{J}_0 - (\mathbf{J}_0 \cdot \nabla)\mathbf{b} + (\mathbf{B}_0 \cdot \nabla)\mathbf{j} - (\mathbf{j} \cdot \nabla)\mathbf{B}_0 + \nu \nabla^2 \mathbf{w}$

The first-order component of (A2) is

A8) $(\mathbf{V}_0 \cdot \nabla)\mathbf{b} + (\mathbf{v} \cdot \nabla)\mathbf{B}_0 - (\mathbf{B}_0 \cdot \nabla)\mathbf{v} - (\mathbf{b} \cdot \nabla)\mathbf{V}_0 = \dfrac{\eta}{\mu_0} \nabla^2 \mathbf{b}$

The equations taken into consideration are the z-component of (A7)

A9) $\quad k\left[B_y''\psi - B_y\nabla^2\psi\right] = \mu_0\rho k\left[V_y''\phi - V_y\nabla^2\phi\right] - i\mu_0\nu\nabla^2\nabla^2\phi$

and the x-component of (A8)

A10) $\quad ik\left[V_y\psi - B_y\phi\right] = \dfrac{\eta}{\mu_0}\nabla^2\psi$

Note that the l.h.s. of (A9) corresponds (apart from pressure and terms related to the geometry ) to the terms appearing in force balance equation (17) with the identifications $F^{m,n} \to kB_y$, $G^{m,n}\sigma' \to kB_y''$.

It is convenient to introduce the following normalization:

A11) $\quad \hat{x} = x/a;\quad \hat{k} = ka;\quad \hat{B}_y = B_y/B_0;\quad \hat{V}_y = V_y/V_0;\quad \hat{\nabla}^2 = \dfrac{d^2}{d\hat{x}^2} - \hat{k}^2$

All these hatted quantities are O(1). Moreover we define

A12) $\quad \hat{\psi} = \psi/aB_0;\quad \hat{\phi} = \phi/aV_0;\quad \hat{\phi}/\hat{\psi} = O(1);$

and the following characteristics times

A13) $\quad \tau_\nu = \dfrac{\rho ka^3}{\nu};\quad \tau_R = \dfrac{\mu_0 ka^3}{\eta};\quad \tau_A^2 = \dfrac{\mu_0\rho a^2}{B_0^2};\quad \tau_{tr} = \dfrac{a}{V_0};$

Therefore (A9) and (A10) become

A14) $\quad \hat{B}_y''\hat{\psi} - \hat{B}_y\hat{\nabla}^2\hat{\psi} = \xi_2\left[\hat{V}_y''\hat{\phi} - \hat{V}_y\hat{\nabla}^2\hat{\phi}\right] - i\xi_3\hat{\nabla}^2\hat{\nabla}^2\hat{\phi}$

A15) $\quad i\left[\hat{V}_y\hat{\psi} - \hat{B}_y\hat{\phi}\right] = \dfrac{1}{\xi_1}\hat{\nabla}^2\hat{\psi}$

where $\quad \xi_1 = \dfrac{\tau_R}{\tau_{tr}},\quad \xi_2 = \left(\dfrac{\tau_A}{\tau_{tr}}\right)^2,\quad \xi_3 = \dfrac{\tau_A^2}{\tau_\nu\tau_{tr}}.$

For the typical plasma parameters we have $\xi_1 \gg 1$, $\xi_2 \ll 1$, $\xi_3 \ll 1$: in fact the characteristic RFX values are $\tau_R \approx \tau_v \approx 0.1s$, $\tau_A \approx 10^{-7}s$, $\tau_{tr} \approx 10^{-4}s$ ($a \approx 0.5m$, $V \approx 5 \cdot 10^3 m/s$) [29]; therefore $\xi_1 \approx 10^3$, $\xi_2 \approx 10^{-6}$, $\xi_3 \approx 10^{-9}$. The smallness of the coefficients $\xi_2$, $\xi_3$ means that the inertial and viscous terms in the force balance equation A14) can be discarded in most of the plasma.

Let's assume the existence of a rational surface, that is a plane $x=x_s$ where $B_y(x_s)=0$. The mode is stationary in the laboratory frame, so according to the no-slip condition [19] we set $V_y(x_s)=0$. In the proximity of this surface we use the following expansions:

$$\hat{B}_y = B'_s \delta + o(\delta^2); \quad \hat{B}''_y = B''_s + B'''_s \delta + o(\delta^2);$$

$$\hat{V}_y = V'_s \delta + o(\delta^2); \quad \hat{V}''_y = V''_s + V'''_s \delta + o(\delta^2);$$

$$\hat{\psi} = \psi_s \left[ \sum_{n=0}^{+\infty} a_n \delta^n + \left( \sum_{n=2}^{+\infty} b_n \delta^n \right) \ln|\delta| \right]; \quad a_0 = 1;$$

$$\hat{\phi} = \phi_s \left[ \sum_{n=0}^{+\infty} \alpha_n \delta^n + \left( \sum_{n=2}^{+\infty} \beta_n \delta^n \right) \ln|\delta| \right];$$

where the subscript "s" denote the value at the rational surface and $\delta = \hat{x} - \hat{x}_s$.

Note that in order to avoid singularities in the $v_y = -\phi'$, $b_y = -\psi'$ perturbations, we set $b_1 = \beta_1 = 0$.

Inserting the previous expansions in (A14), to the leading order one gets $\beta_2 = \beta_3 = \beta_4 = 0$ and the following relation:

$$\text{A16)} \quad B''_s = \frac{\phi_s}{\psi_s} \left\{ \xi_2 V''_s - i\xi_3 \left( \hat{k}^4 - 4\hat{k}^2 \alpha_2 + 24\alpha_4 \right) \right\}.$$

Inserting the expansions in (A15), to the leading order we get $b_2 = 0$, and $2a_2 = \hat{k}^2$.

In a condition of saturated modes it is reasonable to assume that the scale length over which the quantities related to the velocity perturbation change significantly near the rational surface is not smaller than the island width $w$. Taking $\hat{w} = |\hat{\psi}_s|^{1/2}$, this implies the following estimates (A17) $\quad 2|\alpha_2| = \left|\dfrac{1}{\phi_s}\dfrac{d^2\hat\phi}{d\hat x^2}\right|_s \le \dfrac{1}{\hat w^2} = \dfrac{1}{|\hat\psi_s|}; \quad 24|\alpha_4| = \left|\dfrac{1}{\phi_s}\dfrac{d^4\hat\phi}{d\hat x^4}\right|_s \le \dfrac{1}{\hat w^4} = \dfrac{1}{|\hat\psi_s|^2}$

For the typical mode amplitude in RFX we have $|\hat\psi_s| > 10^{-3} \;\to\; \hat w > 10^{-3/2}$. Therefore according to the previous estimate of the parameters $\xi_2$, $\xi_3$, the r.h.s. of (A16) should be very small. We have shown that using the equilibrium system (2, 3) the gradient of the zeroth-order current appearing in the force balance equation must be zero at the rational surface. Here we recover a less strict version of this rule, i.e. the current gradient at the rational surface is not zero but very small

(A18) $\quad B''_s = \dfrac{a^2 \mu_0}{B_0} \left.\dfrac{dJ_{0z}}{dx}\right|_{xs} \ll 1$

The variation of $d\psi/dx$ across the rational surface is given by the radial integral of (A15) over the island region $[x_s-w,\; x_s+w]$. Using the expansions and the condition (A18) one gets:

(A19) $\quad \mathrm{Re}\left(\dfrac{1}{\psi_s}\dfrac{d\hat\psi}{d\hat x}\bigg|_{-\hat w}^{\hat w}\right) = 2\hat k^2 \hat w + \dfrac{2}{3}\xi_1 \hat w^3 \left[-V'_s \,\mathrm{Im}(a_1) + B'_s \,\mathrm{Im}\left(\dfrac{\alpha_1 \phi_s}{\psi_s}\right)\right] + \dfrac{\hat k^4 \hat w^3}{3} + o(\hat w^5)$

The quantity in the l.h.s. of (A19) is the analogous of the parameter $E^{m,n}$, defined by (65). If $|\alpha_1| = \left|\dfrac{1}{\phi_s}\dfrac{d\hat\phi}{d\hat x}\right|_s \sim \dfrac{1}{\hat w}$, the third term of the r.h.s. is of order $\xi_1 \hat w^2$, which is a O(1) quantity for the typical RFX values. Therefore we can have a finite variation of $d\psi/dx$ over the island region. This corresponds to the result of the system (2, 3) where $d\psi/dx$ is discontinuous across the rational surface. Of course this analysis can not predict the sign of this variation.

### Acknowledgments
The authors wish to thank E. Lazzaro, and S. Ortolani for useful discussions.